\documentstyle[twoside,fleqn,espcrc2]{article}

\newcommand{\AmS}{{\protect\the\textfont2
  A\kern-.167em\lower.5ex\hbox{M}\kern-.125emS}}

\title{Anisotropic improved actions}
        
\author{S. Sakai,\address{Faculty of Education, Yamagata University\\},
     A. Nakamura,\address{Research Center for Information Science 
       and Education, Hiroshima University\\}, and
T.Saito,\address{Department of Physics, Hiroshima University}
       }

\begin{document}

\begin{abstract}
  The studies of the quantum corrections for the anisotropy parameter,
$\eta(=\xi_R/\xi_B)$, for the improved actions, 
$\beta (C_0 L(\mbox{Plaq.}) + C_1 L(\mbox{Rect.}))$,
are proceeded in the medium to strong coupling region on anisotropic
lattices.
The global features for the $\eta$ parameters
as a function of $\beta$ and the coefficient $C_{1}$
have been clarified.
It has been found by the perturbative analysis that as $C_1$ decreases,
the slope of the $\eta(\beta)$ becomes less steep and 
for the actions whose $C_{1}$ is less than $-0.160$, $\eta$ decreases as    
$\beta$ decreases, contrary to the case of the standard action.
In the medium to strong coupling region, the $\eta$ parameter begins to 
increase as $\beta$ decreases for all $C_{1}$.
This means that for the  actions with $C_{1} < -0.160$, the one-loop
perturbative results for $\eta$ break down qualitatively and 
the $\eta$ parameters have a dip.
As a result of this dip structure the $\eta$ for Iwasaki's action
remains close to unity in the wide range of $\beta$, which means that
quantum corrections for the anisotropy parameters are small for it.
\end{abstract}

\maketitle

\section{INTRODUCTION}

The advantage of using improved actions has been recognized 
in many area of lattice simulations;  at the same time the merits of
using anisotropic lattice have been well understood\cite{Karsch}.  Then
we had started the study of the anisotropic properties for the improved 
actions\cite{sakaic}.
The improved actions we study in this work consist of terms
$$S_{\mu,\nu}=\beta (C_0 L(\mbox{Plaq.})_{\mu,\nu} + 
C_1 L(\mbox{Rect.})_{\mu,\nu})$$

where $L(\mbox{Plaq.})$ and $ L(\mbox{Rect.})$ represent plaquette and 6-link
rectangular loops respectively, and $C_{0}$ and $C_{1}$
satisfies $C_{0}+8 \cdot C_{1}=1$.
Our improved action covers\\ 
i) tree level Symanzik's action($C_{1}=-1/12$)\cite{Symanzik},\\
ii) Iwasaki's action($C_{1}=-0.331$)\cite{Iwasaki},\\
iii) QCDTARO's action($C_{1}=-1.409$)\cite{Taro}\\
etc.\\
\indent
For these classes of improved actions the anisotropic lattice is formulated
in the same way as in the case of standard plaquette action\cite{Karsch}.
We introduce the coupling constant, $g_{\sigma}$, ($g_{\tau}$) and lattice 
spacing, $a_{\sigma}$, ($a_{\tau}$) in space (temperature) direction. 
With these parameters, the action on the anisotropic lattice is
written as,
$$ S = \beta_{\sigma} \cdot S_{ij} + \beta_{\tau} \cdot S_{4i}, $$
where
$\beta_{\sigma}=g_{\sigma}^{-2} \xi_{R}^{-1}$,
$\beta_{\tau}=g_{\tau}^{-2} \xi_{R}$ and 
$\xi_{R}=a_{\sigma}/a_{\tau}$.\\  
\indent
In the weak coupling expansions, the $\eta$ parameter 
has been written as follows,
$$\eta = 1+N \cdot \alpha(\xi_{R},C_{1})/\beta
+O(g^{4}). $$ 
The behavior of $\alpha(\xi_{R},C_{1})$ had already been reported
at lat98\cite{sakaic}, which we will show in Fig.1.
\begin{figure}[htb]
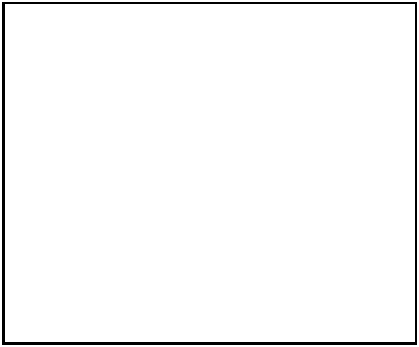

\vspace{9pt} 
\framebox[55mm]{\rule[-21mm]{0mm}{43mm}}
\caption{The coefficients $\alpha$ as a function of
$C_{1}$}
\label{fig:largenenough}
\end{figure}
We have found that $\alpha(\xi_{R},C_{1})$ changes sign around 
$C_{1}\sim -0.160$.  Namely in the weak coupling region, for the
Iwasaki's and QCDTARO's actions,
the $\eta$ parameter is less than unity contrary to the case of
the standard plaquette action.\\
\indent
The natural question is what is the behavior of $\eta$ in the
intermediate to weak coupling region, for those improved actions.

\section{NUMERICAL STUDY of $\eta$}

Numerically, the parameter $\eta$ is calculated by the
relation $\eta = \xi_{R}/\xi_{B}$\cite{Klassen}\cite{Nakamura}\cite{Engels}
, where $\xi_{B}$ is a bare
anisotropy parameter which appears in the action,
$$ S = \beta(\frac{1}{\xi_{B}}S_{ij}+\xi_{B}S_{i4}), $$
while the renormalized anisotropy parameter is defined by 
$\xi_{R}= a_{\sigma}/a_{\tau}$.  For the probe of the scale for the
space and temperature direction, we use the lattice potential 
in these directions. The lattice potential in temperature direction 
is defined by Wilson loops in space-temperature plane,
$$V_{st}(\xi_{B},p,t)=\ln(\frac{W_{st}(p,t)}{W_{st}(p+1,t)}).$$
The potential in the space direction is defined in a similar way.\\
\indent
The simulations are mainly done on the $12^{3} \times 24 $ lattice
for $\xi_{R}=2.0$,
and for some $\beta$ on $16^{3} \times 32$ lattice to study the size 
dependences.\\
\indent
The method of determination of $\eta$ is explained in some detail
by the examples of $\beta=4.5$ for Iwasaki's improved action. \\
\indent
We fix $\xi_{R}=2$, and calculate the rate 
$$R(\xi_{B},p,r)=\frac{V_{ss}(\xi_{B},p,r)}{V_{st}(\xi_{B},p,\xi_{R}t)}$$
at $\xi_{B}=2.0,2.1,2,2$. The results for $\xi_{B}=2.1$ are shown
in the Fig.2 for each $p$ and $r$.
\begin{figure}[htb]
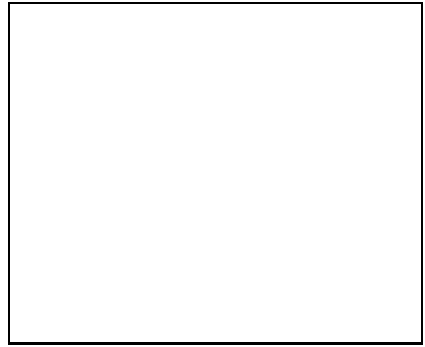

\vspace{9pt} 
\framebox[55mm]{\rule[-21mm]{0mm}{43mm}}
\caption{Ratio $R(p,r)$ for $\beta=4.5$ and $\xi_{B}=2.1$}
\label{fig:largenenough}
\end{figure}
For these $R(\xi_B)$ we have made the interpolation
by the second order polynomial in $\xi_{B}$ and solve the point where
$R=1$ is satisfied.
Then the $\eta$ is obtained for each $p$ and $r$. 
The results for the $\eta$ are shown in Fig.3.
\begin{figure}[htb]
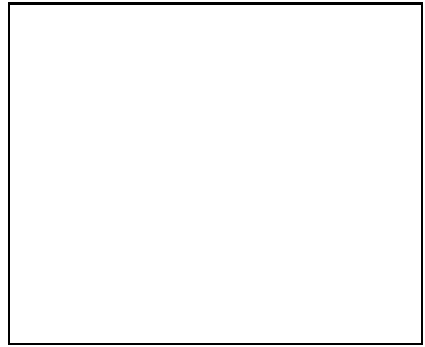

\vspace{9pt} 
\framebox[55mm]{\rule[-21mm]{0mm}{43mm}}
\caption{$\eta(p \times r)$ at $\beta=4.5$ and $\xi_{R}=2.0$}
\label{fig:largenenough}
\end{figure}
It is found that for $p \times r \ge 15$, $\eta$ is almost flat. 
For the flat region, we take
the average of them. The errors are estimated by the Jackknife method.
The same analysis has been repeated for other $\beta$ and for the other
Improved actions.\\
\indent
In order to check the size dependences, the simulations are carried out on
$16^3 \times 32$ lattice at $\beta=21.0$ and $3.5$ for the Iwasaki's 
improved action.  It is found that at
$\beta=21.0$ the size dependence is not small but at $\beta=3.5$ it is
quite small. Therefore for the intermediate to strong coupling region,
the simulation is carried out on $12^3 \times 24$ lattice.
All the results are summarized in Fig.4.

\section{CONCLUSIONS AND DISCUSSIONS}
\begin{figure}[htb]
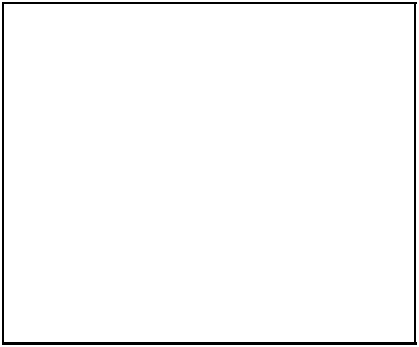

\vspace{9pt} 
\framebox[55mm]{\rule[-21mm]{0mm}{43mm}}
\caption{All the results for $\eta(\beta,C_{1})$. 
The results for the standard action is from Ref.6}
\label{fig:largenenough}
\end{figure}
$\bullet$ Iwasaki's Action\\
\indent
The $\eta$ parameter has a shallow dip at $\beta \sim 6.0$ and 
then blows up towards the small $\beta$ region. The prediction by the
one loop perturbative calculation breaks down qualitatively.  
As a results of this behavior
the $\eta \sim 1$ in the wide range of $\beta$.\\ 
\indent
The results for $\xi_{R}=3$ are also shown in the Fig.4. It is found that
the qualitative behavior is the same but the dip around $\beta=6.0$
becomes deeper. This means that the renormalization of the anisotropy
parameter is larger for larger $\xi_{R}$.\\

$\bullet$ Tree-level Symanzik's Action and the action with $C_{1}=-0.16$\\
\indent
The $\eta$ for Symanzik's action is qualitatively the same as those
of the standard action\cite{Engels}; namely it increases monotonically
as $\beta$ decreases. But the slope is 
less steep and then the value of  $\eta$ is smaller than 
the standard action at a given $\beta$.\\
\indent
For the action with $C_{1}=-0.16$, it has been found that $\alpha=0$, 
namely the
renormalization of the anisotropy parameter is zero in the weak
coupling perturbative calculations\cite{sakaic}. 
The $\eta$ in the intermediate to
strong coupling region shows similar behavior with
the standard plaquette action and Symanzik's action. 
But again the slope becomes less steep.\\

$\bullet$ Global features of $\eta(\beta,C_{1})$\\ 
\indent
The calculation with the QCDTARO's action is not carried out yet. But we
have found the global structure of the $\eta$ parameter as a function of 
$\beta$ and $C_{1}$. The interesting point is that the for the action
with $C_{1} < -0.16$, the $\eta$ once decreases and then increases with
decreasing $\beta$. As a result of this behavior the $\eta$ stays close 
to one in the wide range of $\beta$ for Iwasaki's improved action. 
This may be a good features for the numerical simulations.\\

$\bullet$  Further works\\
\indent
     The calculation of the $\eta$ at stronger coupling region and
the other $\xi_{R}=0.5, 1.5, 3.0
5.0$ etc are now under study.  We are preparing to use the smeared
Wilson loops for the simulation in the stronger coupling region.\\ 
\indent
We are starting the calculation of the lattice spacing $a$ as a function
of $\beta$ on the anisotropic lattices.\\
\indent
    The behavior of $\eta$ with smaller
$C_{1}$ like QCDTARO's action, is also an interesting problem. \\
\indent
  We are preparing the simulations of physical 
quantities on an anisotropic lattice with improved actions. 
The targets are 
heavy quark spectroscopy, transport coefficients of quark
gluon plasma etc. \\

ACKNOWLEDGMENTS\\
The present work has been done with SX-4 at RCNP and VX-4 at
Yamagata. We are grateful for the members of RCNP for warm hospitality
and kind supports.\\

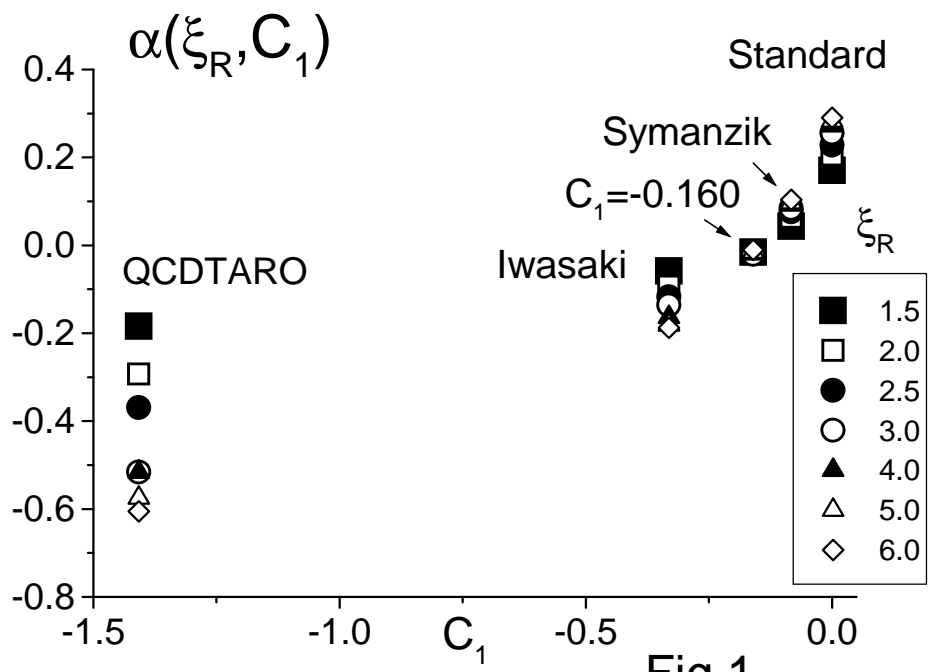
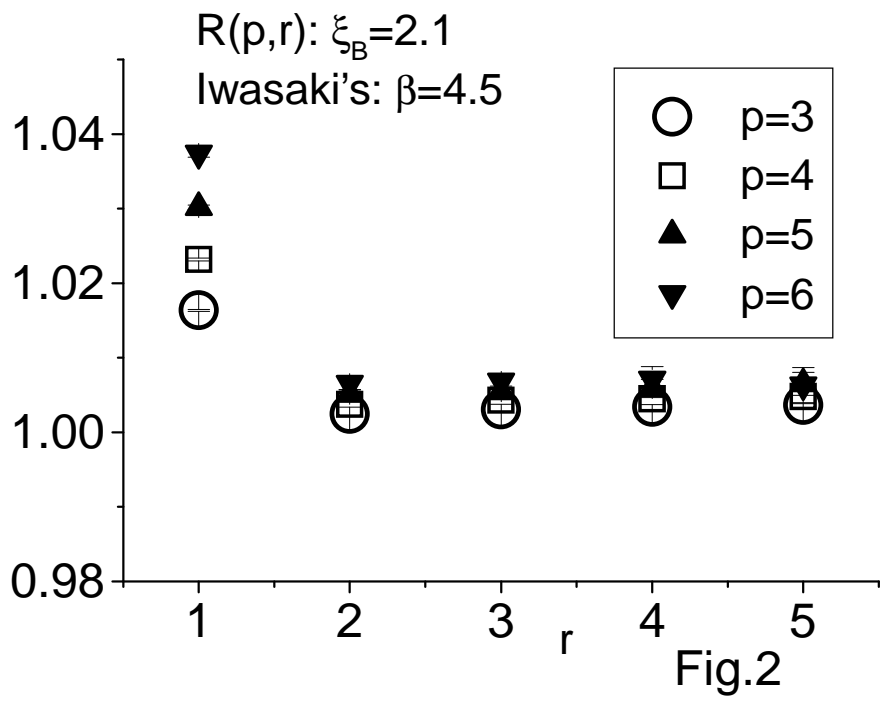
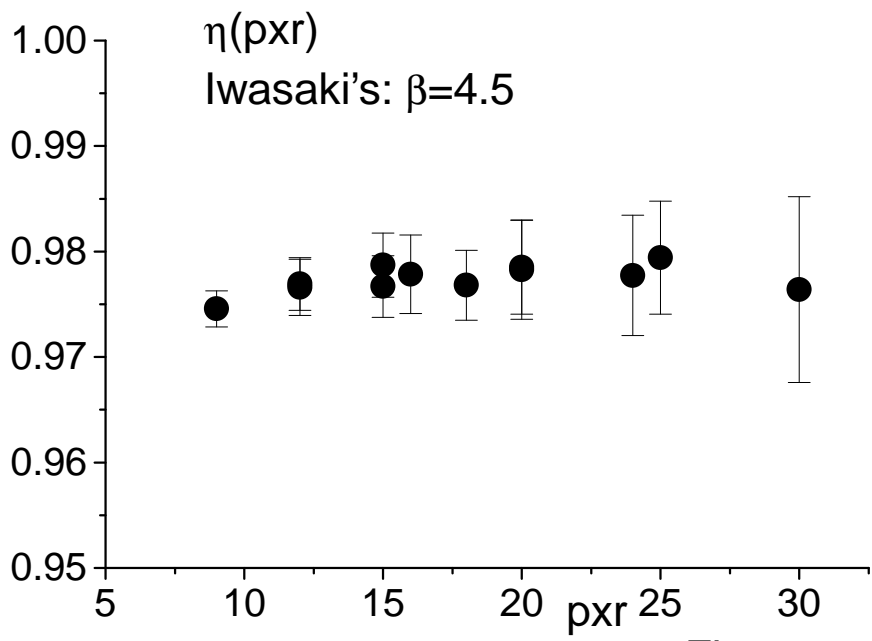
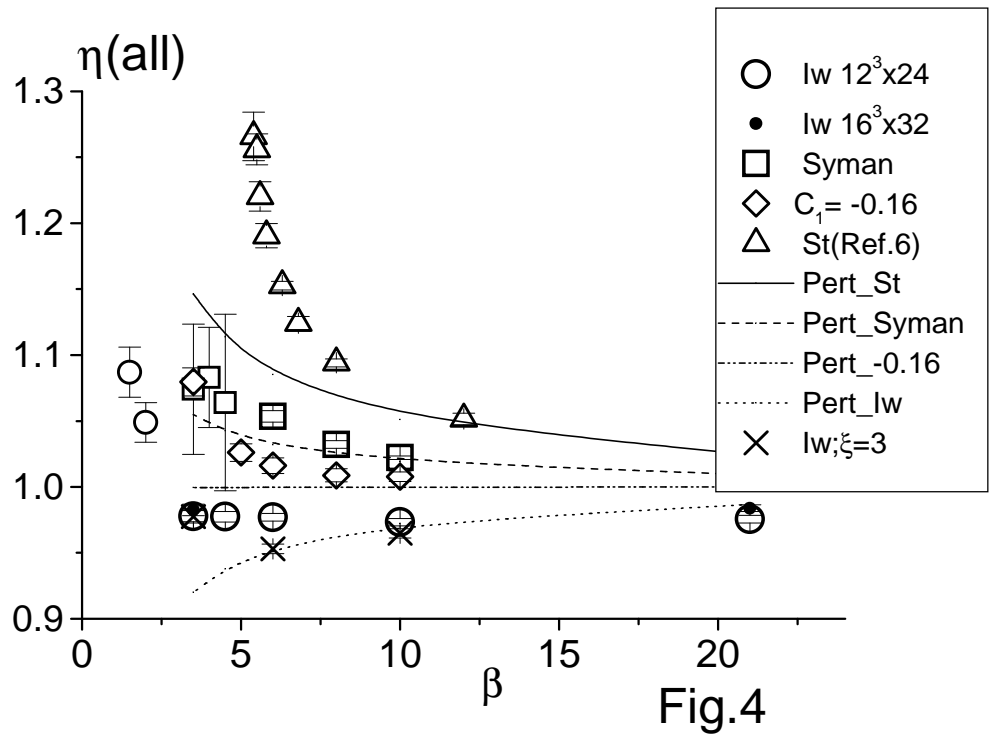
\end{document}